\documentclass[aps,prl,superscriptaddress, twocolumn, showpacs]{revtex4-1}
\usepackage{graphicx}
\usepackage{dcolumn}
\usepackage{bm}
\usepackage{amsmath}

\usepackage{color}
\usepackage{bm,graphicx,hyperref}
\hypersetup{%
  breaklinks = {true},
  citecolor = {blue},
  colorlinks = {true},
  linkcolor = {red},
}

\begin{document}

\title{Accelerated search and design of stretchable graphene kirigami
  using machine learning}

\author{Paul~Z.~Hanakata}
\affiliation{Department of Physics, Boston University, Boston, MA 
02215}
\email{paul.hanakata@gmail.com}

\author{Ekin~D.~Cubuk} \affiliation{Google Brain, Mountain View, CA 94043, USA}

\author{David~K.~Campbell}
\affiliation{Department of Physics, Boston University, Boston, MA  02215}

\author{Harold~S.~Park}
\affiliation{Department of Mechanical Engineering, Boston University, Boston, MA 
02215}

\date{\today}
\begin{abstract}
  Making kirigami-inspired cuts into a sheet has been shown to be an
  effective way of designing stretchable materials with metamorphic
  properties where the 2D shape can transform into complex 3D
  shapes. However, finding the optimal solutions is not
  straightforward as the number of possible cutting patterns grows
  exponentially with system size. Here, we report on how machine
  learning (ML) can be used to approximate the target properties, such
  as yield stress and yield strain, as a function of cutting
  pattern. Our approach enables the rapid discovery of kirigami
  designs that yield extreme stretchability as verified by molecular
  dynamics (MD) simulations. We find that convolutional neural
  networks (CNN), commonly used for classification in vision tasks,
  can be applied for regression to achieve an accuracy close to the
  precision of the MD simulations. This approach can then be used to
  search for optimal designs that maximize elastic stretchability with
  only 1000 training samples in a large design space of
  $\sim 4\times10^6$ candidate designs.  This example demonstrates the
  power and potential of ML in finding optimal kirigami designs at a
  fraction of iterations that would be required of a purely MD or
  experiment-based approach, where no prior knowledge of the governing
  physics is known or available.
\end{abstract}

\maketitle
\emph{Introduction--}Recently, there has been significant interest in
designing flat sheets with metamaterial-type properties, which rely
upon the transformation of the original 2D sheet into a complex 3D
shape.  These complex designs are often achieved by folding the sheet,
called the origami approach, or by patterning the sheet with cuts,
called the kirigami approach. Owing to the metamorphic nature, designs
based on origami and kirigami have been used for many applications
across length scales, ranging from meter-size deployable space
satellite structures~\cite{zirbel-JMD-135-111005-2013} to soft
actuator crawling robots~\cite{rafsanjani-SR-3-7555-2018} and
micrometer-size stretchable electronics~\cite{shyu-NatMat-14-785-2015,
  blees-Nature-524-204-2015}.

Atomically thin two-dimensional (2D) materials such as graphene and
MoS$_2$ have been studied extensively due to their exceptional
physical properties (mechanical strength, electrical and thermal
conductivity, etc). Based on
experiments~\cite{blees-Nature-524-204-2015} and atomistic
simulations~\cite{zenan-PRB-90-245437-2014,
  hanakata-Nanoscale-8-458-2016}, it has been shown that introducing
arrays of kirigami cuts allows graphene and MoS$_2$ to buckle in the
direction perpendicular to the plane. These out-of-plane buckling and
rotational deformations are key to enabling significant increases in
stretchability.

By the principles of mechanics of springs, it is expected that adding
cuts (removing atoms) generally will both soften and weaken the
material. Griffith's criterion for fracture~\cite{griffith-1921} has
been successfully used to explain the decrease in fracture strength
for a single cut~\cite{zhao-JAP-108-064321-2010,
  zhang-NatComm-5-3782-2014, jung-EML-2-52-2015,
  rakib-physicaB-515-67-2017}, but cannot explain how the delay of
failure is connected to the out-of-plane deflection of kirigami
cuts. Several analytical solutions have been developed to explain the
buckling mechanism in single cut
geometries~\cite{dias-sm-48-9087-2017, isobe-SR-6-24758-2016}, a
square array of mutually orthogonal
cuts~\cite{rafsanjani-PRL-118-084301-2017}, and a square
hole~\cite{moshe2018nonlinear}. These analytical solutions are
applicable for regular repeating cuts, but may not be generally
applicable for situations where non-uniform and non-symmetric
cuts may enable superior performance.

An important, but unresolved question with regards to kirigami
structures at all length scales is how to locate the cuts to achieve a
specific performance metric.  This problem is challenging to
solve due to the large numbers of possible cut configurations that
must be explored.  For example, the typical size scale of current
electronic devices is micrometers ($10^{-6}$ meters) and the smallest
cuts in current 2D experiments are about
$10\times10$~\AA~\cite{masih-ACS-10-5687-2016}. Thus, exhaustively
searching for good solutions in this design space would be impractical
as the number of possible configurations grows exponentially with the
system size.  Alternatively, various optimization algorithms,
i.e. genetic and greedy algorithms, and topology optimization
approaches, have been used to find optimal designs of materials based
on finite element methods~\cite{sigmund-16-68-1998,
  jakiela-elsevier-186-2-339-2000, huang-43-1039-2007,
  gu-JAM-83-071006-2016}.  However, these approaches have difficulties
as the number of degrees of freedom in the problem increases, and also
if the property of interest lies within the regime of nonlinear
material behavior.

Machine learning (ML) methods represent an alternative, and recently
emerging approach to designing materials where the design space is
extremely large.  For example, ML has been used to design materials
with low thermal conductivity~\cite{seko-PRL-115-205901-2015}, battery
materials~\cite{sendek2017holistic,onat2018implanted}, and composite
materials with stiff and soft components~\cite{gu-EML-18-19-2018}.  ML
methods have also recently been used to study condensed matter systems
with quantum mechanical
interactions~\cite{behler2007generalized,behler2011atom,cubuk2017representations},
disordered atomic
configurations~\cite{cubuk2016structural,schoenholz2017relationship,cubuk2017structure}
and phase
transitions~\cite{carrasquilla-NatPhys-13-431-2017,broecker2017machine}. While
ML is now being widely used to predict properties of new materials,
there have been relatively few demonstrations of using ML to design
functional materials and structures~\cite{gu-MatHor-2018}.

\begin{figure}
\includegraphics[width=6cm]{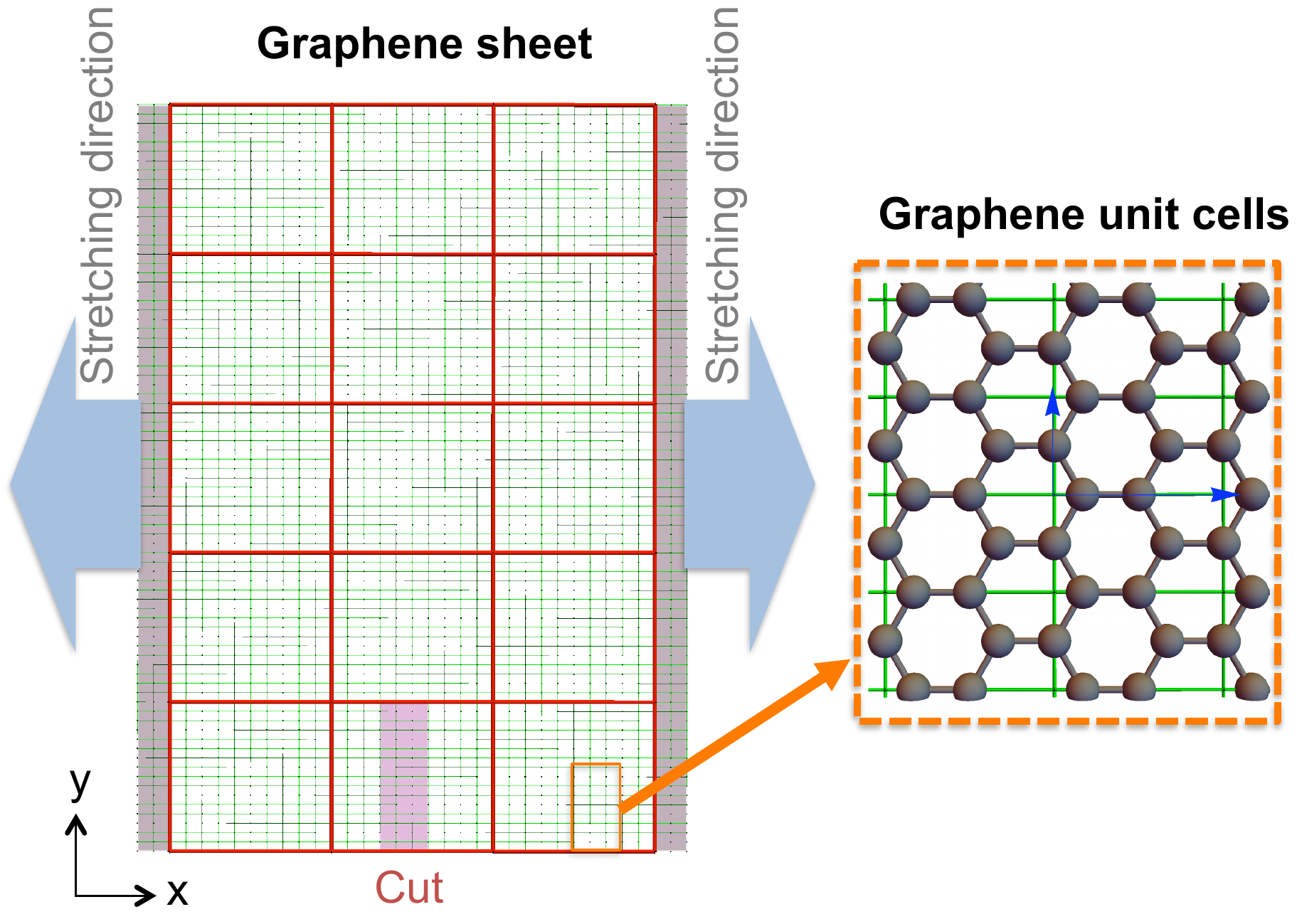}
\caption{Schematic diagrams of graphene sheet and rectangular graphene
  unit cells. Each of the grid (colored red) consists of $10\times16$
  rectangular graphene unit cells (colored green). }
\label{fig:fig1}
\end{figure}

In this letter, we use ML to systematically study how the cut density
and the locations of the cuts govern the mechanical properties of
graphene kirigami. We use fully-connected neural networks (NN) and
also convolutional neural networks (CNN) to approximate the yield
strain and stress.  To formulate this problem systematically, we
partition the graphene sheets into grids, where atoms in each grid
region will either be present or cut, as shown schematically in
fig.~\ref{fig:fig1}. We then utilize the CNN for inverse design,
where the objective is to maximize the elastic stretchability of the
graphene kirigami subject to a constraint on the number of cuts. We
use ML to search through a design space of approximately 4,000,000
possible configurations, where it is not feasible to simulate all
possible configurations in a brute force fashion. Despite the size of
the design space, our model is able to find the optimal solution with
fewer than 1000 training data points (evaluations via MD). Our
findings can be used as a general method to design a material without
any prior knowledge of the fundamental physics, which is particularly
important for designing materials when only experimental data are
available and an accurate physical model is unknown.

\emph{Overview of mechanical properties--} 
In this section, we give a
brief overview of the changes in the mechanical properties of graphene with
cuts.  The 2D binary array of cut configurations
$N_{\rm grid}=N_x\times N_y$ is flattened into a one-dimensional array
vector $\pmb{x}$ of size $n=N_{\rm grid}$. We use $n$ for number of
features, $m$ for the number of samples,
$\pmb{x}=(x_1, x_2, ..., x_{n})^T$ for the binary vector describing
cut configurations, $\vec{x}, \vec{y}, \vec{z}$ for the real space
vectors (atomic locations), and $\hat{x}, \hat{y}, \hat{z}$ for the
unit vectors in real space.

We study one unit kirigami of size $\sim100\times200$~\AA, where cuts
are allowed to be present on the $3\times5$ grid; this gives a design
space of $2^{15}=32768$ possible cut configurations
(fig.~\ref{fig:fig1}). Each cell of the grid also consists of
$10\times16$ rectangular graphene unit cells.  There are 2400
rectangular graphene unit cells in this sheet; there are four carbon
atoms in the rectangular graphene unit cell. This gives a total of
9600 carbon atoms in a kirigami sheet without cuts.  In this system,
the cut density can range from 0 cuts in the 15 grids to 15 cuts in
the 15 grids, while keeping each cut size constant at
$12\times38$~\AA~($3\times16$ rectangular graphene unit cells), which
is relevant to current experimental
capabilities~\cite{masih-ACS-10-5687-2016}.  Following previous
work~\cite{zenan-PRB-90-245437-2014, dias-sm-48-9087-2017}, we use the
Sandia open-source MD simulation code LAMMPS (Large-scale
Atomic/Molecular Massively Parallel Simulator)~\cite{plimptonLAMMPS}
to generate the ground truth data for our training model, where we
simulate graphene as the 2D constituent material of choice for the
kirigami at a low temperature of $4.2$ K.  Since we simulate MD at
$T=4.2$ K, the obtained yield strain (or stress) of a configuration
varies due to stochasticity (i.e. distributions of the initial
velocities). The MD precisions for strain and stress are
$\eta^{\varepsilon}=0.046$ and $\eta^{\sigma}=2.00$ GPa,
respectively. In this work, we focus only on kirigami with armchair
edges along the $\hat{x}$ direction as the stretchability is improved
regardless of the chirality of graphene with armchair or zigzag
edges~\cite{zenan-PRB-90-245437-2014}. The sheets are stretched in the
$\hat{x}$ direction and engineering strain $\varepsilon=L/L_0-1$ is
used to quantify stretchability, where $L_0$ and $L$ are the length of
sheet in the direction of the loading before and after the
deformation, respectively. More details of simulations can be found in
the supplemental material (SM).

\begin{figure}
\includegraphics[width=8cm]{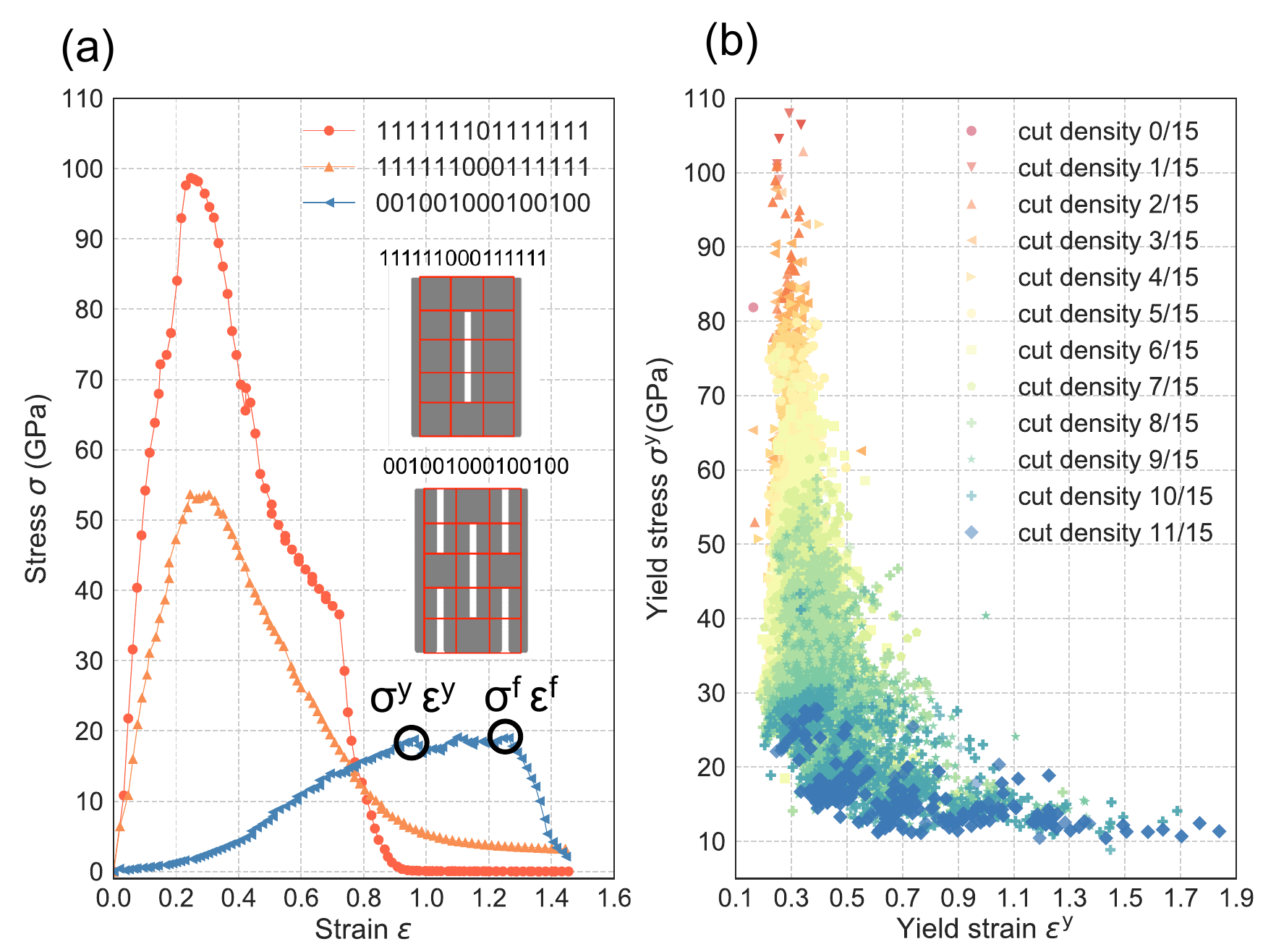}
\caption{ (a) Stress-strain plot of three representative
  kirigamis. Inset shows the ``typical'' kirigami cuts. (b) Yield
  stress as a function of yield strain for different
  configurations. Data are colored based on their cut density. }
\label{fig:fig2}
\end{figure}

Stress-strain curves of three representative cuts are shown in
fig.~\ref{fig:fig2}(a). For the remainder of the paper, we will focus
on the yield point where plastic deformation/bond breaking occurs; the
two quantities of interest are yield stress $\sigma^{\rm y}$ and yield
strain $\varepsilon^{\rm y}$. As shown in fig.~\ref{fig:fig2}(b), the
$\sigma^{\rm y}$ (stress at which bond breaking occurs) consistently
decreases with increasing number of cuts. $\varepsilon^{\rm y}$ has
much more variability at higher cut density. At a cut density of 73\%
(11 cuts), $\varepsilon^{\rm y}$ varies over a wide range of values
from $\sim 0.2$ (20\%) to $\sim 2.0$ (200\%). This shows that
increasing number of cuts without intelligently locating the cuts may
not always increase the stretchability. 

\emph{Machine learning--} We trained NNs and CNNs to predict the yield
strain in the context of supervised learning.  2D images of size
$30\times80$ are used as inputs for training the CNN. For the NN, the
2D images are flattened to 1D arrays of size 2400. The 2400 grids
correspond to the number of rectangular graphene unit cells. In vision
tasks CNN is usually used for classification. Here, we will use both
NN and CNN for {\it regression}. Accordingly, we do not include the
activation function at the end of the final layer, and we minimize the
mean squared error loss to optimize the model parameters.

Since the yield strain and yield stress results are similar as they
are, roughly, inversely proportional to each other (see
fig.~\ref{fig:fig2}(b)), we will focus on the yield strain. All plots
and data for yield stress can be found in the SM. Out of $2^{15}$
possible configurations, only the 29,791 non-detached configurations
are considered. We split the 29,791 data samples into 80\% for
training, 10\% for validation, and 10\% for test dataset. The
validation dataset is used to find better architectures
(``hyperparameter tuning'', i.e. changing number of neurons or
filters), and the test dataset is used to assess performance. We
provide details on the hyperparameters and the performance of
different CNN and NN architectures in the SM.

We use simple shallow NNs with one hidden layer of size ranging from 4
and 2024.  For CNN, we use architectures similar to
VGGNet~\cite{vggNet2014}. The kernel size is fixed at $3\times3$ with
a stride of 1. Each convolutional layer is followed by a rectified
linear unit (ReLU) function and a max-pooling layer of size $2\times2$
with stride of 2~\cite{lecun-IEEE-86-2278-1998}. Based on validation
dataset performance, here we report the best performing CNN and NN
architecture.

We use the root-mean-square error (RMSE) and $R^2$ to evaluate the
goodness of a model.  A CNN with number of filters of 16, 32, 64 in
first, second, and third convolutional layer, respectively, and a
fully-connected layer (FCL) of size 64 achieves $R^2=0.91$ and RMSE of
0.054 which is close to the MD precision of 0.046. We will denote this
CNN model by CNN-f16-f32-f64-h64; here `f' stands for filter and `h'
stands for number of neurons in the FCL. A NN with 64 neurons achieves
$R^2=0.84$ and RMSE of 0.075. A NN with 246 neurons achieves a RMSE of
0.123 and CNN with 256 FCL achieves a RMSE of 0.054. We found that
making NN wider (increasing number of neurons) does not improve the
accuracy. In addition, we use simple ordinary least square (OLS)
regression to see how CNN performs compared to such simpler model.
For yield strain, a polynomial degree of 3 gives $R^2=0.76$ and
${\rm RMSE}=0.084$. The CNN performs better than NN and OLS as the CNN
learns from the local 2D patterns. Performance of CNN and NN with
different architectures (different neurons number ranging from 4
to 2024) as well as simple OLS, and details on RMSE, MD precision and
$R^2$ can be found in the SM.

\begin{figure*}
\includegraphics[width=16cm]{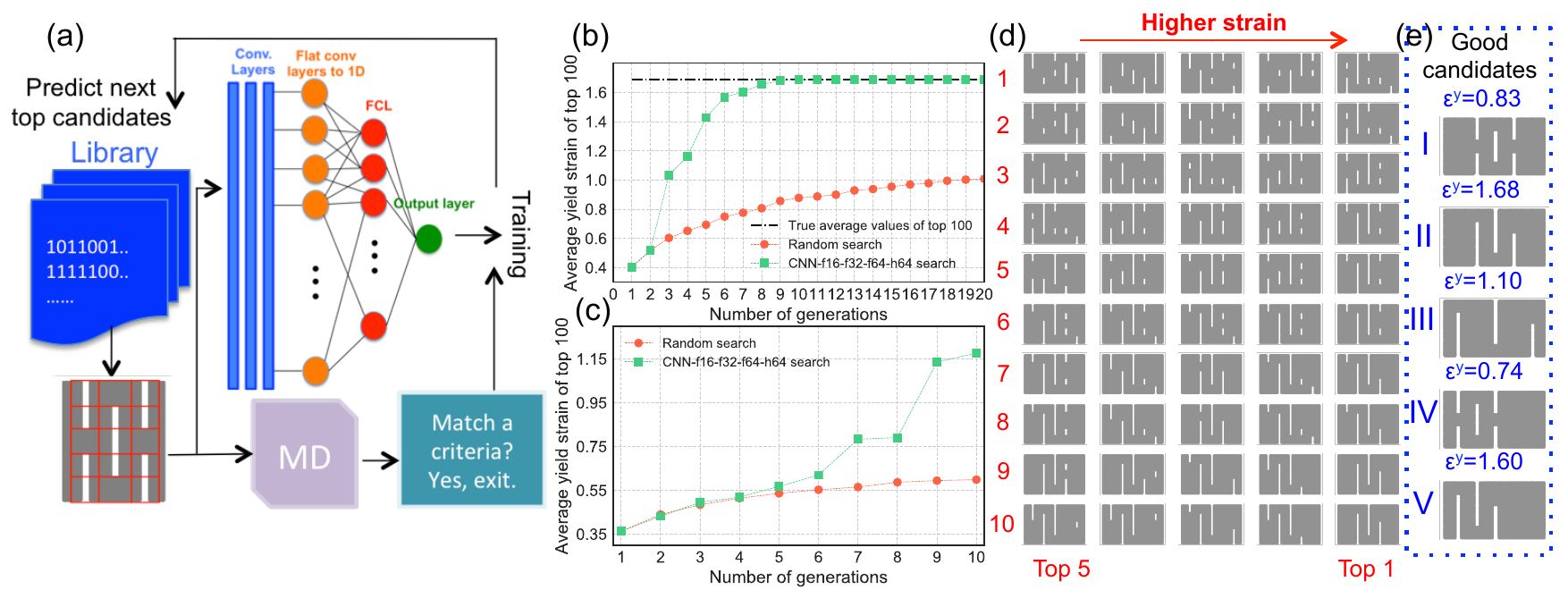}
\caption{(a) Schematic of the neural network search algorithm. Average yield
  strains of the top 100 performers as a function of number of
  generations for kirigami with allowed cuts of (b) $3\times5$ and (c)
  $5\times5$. (d) Visualization of top five performers of kirigami
  with $5\times5$ allowed cuts in each generation. After the eighth
  generation, the top five performers remain constant. (e) A
  comparison between the top performing configurations found by the ML
  and the typical kirigami configurations with centering cuts. Note
  that the kirigami visualizations are not scaled to the real physical
  dimensions for clarity. }
\label{fig:fig3}
\end{figure*}

\emph{Inverse Design of Highly Stretchable Kirigami--} In the previous
section, we used NN and CNN for the prediction of mechanical
properties, in the context of supervised learning. Next, we
investigate if the approximated function can be used to search for
optimal designs effectively. Here, we will use CNN, the best
performing model, to search for the cut configuration with the largest
yield strain. The procedure is as follows: first we randomly choose
100 configurations from the library of all possible configurations and
use MD to obtain the yield strain. After training, the CNN then is
used to screen the {\it unexplored} data set for the top performing
100 remaining candidates. Based on this screening, 100 new MD
simulations are performed and the results are added to the training
set for the next generation. The ML search algorithm flow diagram is
shown in fig.~\ref{fig:fig3}(a). The difference from the previous
section is that here we train the model incrementally with the
predicted top performers.

We first use the $3\times5$ allowed cuts where we already have
simulated {\it all} of the possible configurations in MD to make sure
that our model indeed finds the {\it true} (or close to) optimal
designs. To evaluate the performance of the search algorithm we use
the average of yield strain of the top 100 performers
$\overline{\varepsilon^{\rm y}}_{\rm top 100}$ for each generation.
This number, which cannot be too small, is chosen arbitrarily so that
we obtain more than a handful of good candidates.  As a benchmark, we
include the `naive' random search. Specifically, we use
CNN-f16-f32-f64-h64 architecture to find the optimal designs. As shown
in fig.~\ref{fig:fig3}(b), the random search needs 30 generations
(3000 MD simulations) to get
$\overline{\varepsilon^{\rm y}}_{\rm top 100}\geq1.0~(100\% {\rm
  strain})$
and explore the entire design space in order to find the true best 100
performers. The CNN approach requires only {\it 3 generations} (300 MD
simulations data) to search for 100 candidates with
$\overline{\epsilon^{\rm y}}_{\rm top 100}\sim1.0$ and $10$
generations to search the {\it true} top 100 performers. In each
generation the standard deviation of
$\overline{\epsilon^{\rm y}}_{\rm top 100}$ is around 0.25. Using CNN
to search for optimal designs is crucial because one MD simulation of
graphene with a size of $100\times200$~\AA~requires around 1 hour
computing time using 4 cores of CPU. In each generation, the required
time to train the CNN and to predict the yield strain of one
configuration is around 6 milliseconds on 4 CPU cores (same machines)
or 3 milliseconds on 4 CPU cores plus one GPU~\footnote{Details of the
  specific GPU can be found in the SM}.  From fig.~\ref{fig:fig2}(b),
we know that sheets with high strains are ones with high cut
density. However, the variability is also large; for example at 11/15
cut density, the yield strain ranges from 0.2 to 1.7. Despite of this
complexity, the ML quickly learns to find solutions at high cut
density and also to find the right cutting patterns.

Next, we apply this simple algorithm to a much larger design space
where the true optimal designs are {\it unknown} and also with a
specified design constraint. Specifically, we study larger graphene
sheets by extending the physical size in $\hat{x}$ from
$\sim100\times200$ to $\sim200\times200$~\AA~(from $30\times80$ to
$50\times80$ rectangular graphene unit cells). For this system, one MD
simulation requires around 3 hours of computing time running on 4
cores.  The allowed cuts are also expanded from $3\times5$ to
$5\times5$ grids. For this problem, we fixed number of cuts at 11
cuts, which gives a design space of size
$\frac{25!}{11!14!}\sim 4\times 10^6$.  For this system, we could not
use brute force to simulate all configurations as we did previously
for system with 15 allowed cuts. While the typical stretchable
kirigamis usually have cuts and no-cuts along the loading direction
($\hat{x}$), it is not clear whether all the cuts should be located
closely in a region or distributed equally.

As shown in fig.~\ref{fig:fig3}(b), the CNN is able to find designs
with higher yield strains. With fewer  than 10 generations (1000 training
data), the CNN is able to find configurations with yield strains
$\geq1.0$, which is roughly {\it five} times larger than a sheet
without cuts. In each generation, the standard deviation of the top
100 performers is around 0.1. In fig.~\ref{fig:fig3}(d), we plot cut
configurations of the top five performers in each generation. It can
be seen that the cut configurations are random in the early stage of
the search but evolve quickly to configurations with a long cut
along the $\hat{y}$ direction alternating in $\hat{x}$ direction, as we
expected from the smaller grid system. This suggests that our ML
approach is scalable in a sense that the same CNN architecture used
previously in the simpler system with 15 allowed cuts can search the
optimal designs effectively despite a large design space.

We next take a closer look on the top performing
configurations. Interestingly, the optimal solutions for maximum
stretchability found by CNN have cuts at the edges which are different
from the ``typical'' kirigami with centering cuts
(fig.~\ref{fig:fig3}(e) configuration I). The found best performer has
a yield strain {\it twice} as large as the kirigami with centering
cuts. We found that to achieve high yield strains the long cuts should
be located close to each other, rather than being sparsely or equally
distributed across the sheet along the $\hat{x}$ direction, as shown
by comparing configurations II and III in
fig.~\ref{fig:fig3}(e). These overlapping cut configurations allow
larger rotations and out-of-plane deflection which give higher
stretchability, i.e. the alternating edge cut pattern effectively
transforms the 2D membrane to a quasi-1D membrane. Close packing of
the alternating edge cuts allows increased stretchability because the
thinner ribbons connecting different segments improve twisting. This
result is similar to what we found previously in kirigami with
centering cuts~\cite{zenan-PRB-90-245437-2014,
  hanakata-Nanoscale-8-458-2016}. Visualizations illustrating these
effects and a more detailed discussion can be found in the SM. This
design principle is particularly useful as recently a combination of
dense and sparse cut spacing were used to design stretchable thin
electronic membranes~\cite{hu-PRApplied-9-021002-2018}.  It is
remarkable not only that ML can quickly find the optimal designs using
few training data ($<1\%$ of the design space) under certain
constraints, but also that ML can capture uncommon physical insights
needed to produce the optimal designs, in this case related to the cut
density and locations of the cuts.

\emph{Conclusion--} We have shown how machine learning (ML) methods
can be used to design graphene kirigami, where yield strain and stress
are used as the target properties. We found that CNN with three
convolutional layers followed by one fully-connected layer is
sufficient to find the optimal designs with relatively few training
data. Our work shows not only how to use ML to effectively search for
optimal designs but also to give new understanding on how kirigami
cuts change the mechanical properties of graphene sheets.
Furthermore, the ML method is parameter-free, in a sense that it can
be used to design any material without any prior physical knowledge of
the system.  As the ML method only needs data, it can be applied to
experimental work where the physical model is not known and cannot be
simulated by MD or other simulation methods. Based on previous work
indicating the scale-invariance of kirigami
deformation~\cite{dias-sm-48-9087-2017}, the kirigami structures found
here using ML should also be applicable for designing larger
macroscale kirigami structures.

\begin{acknowledgments}
 
\end{acknowledgments}

\bibliography{ml}
\pagebreak
\begin{center}
\textbf{\large Supplemental Materials}
\end{center}

\section{Molecular dynamics methods}

\begin{figure}
\includegraphics[width=8cm]{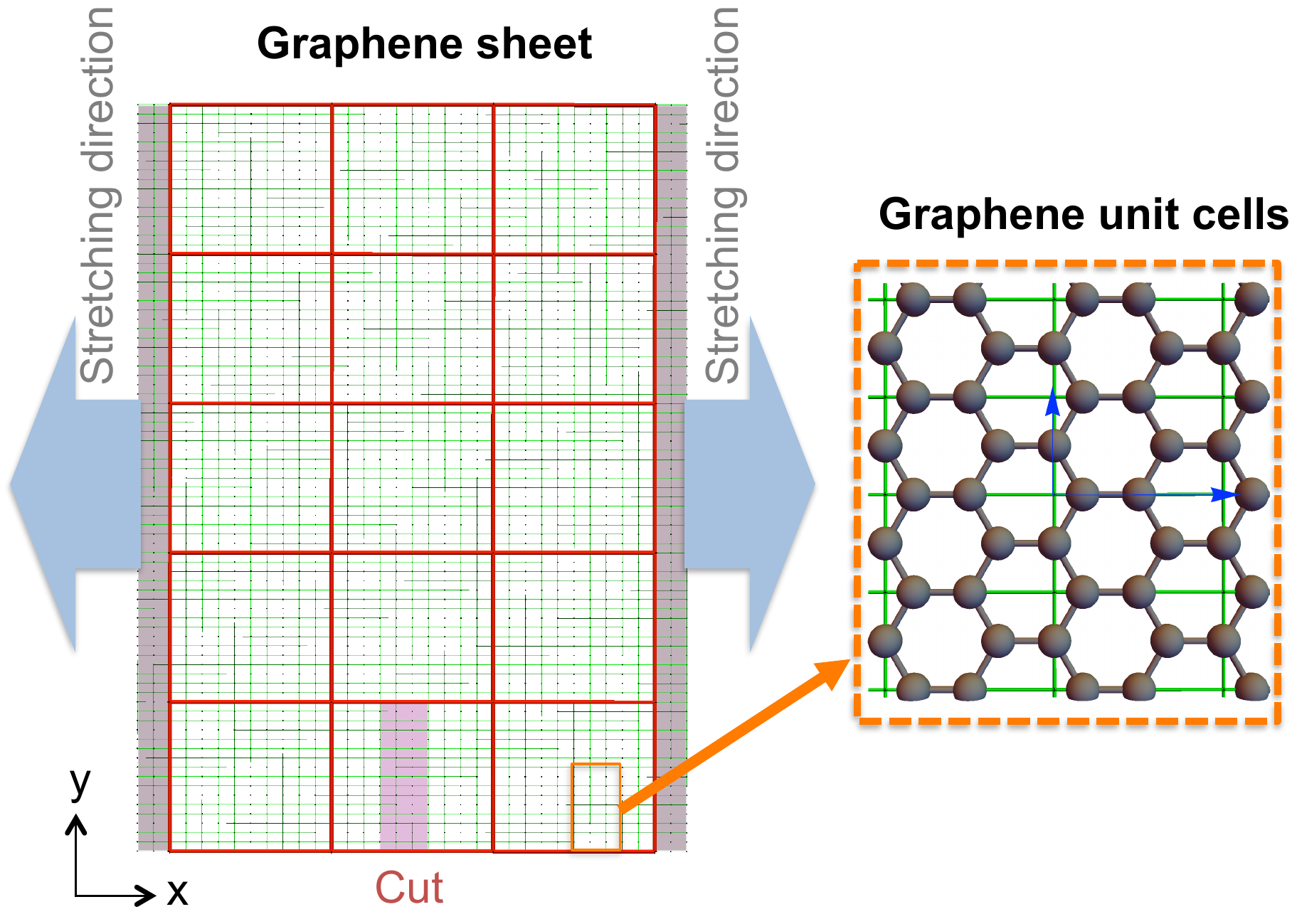}
\caption{Schematic diagrams of graphene sheet and rectangular graphene
  unit cells. In this system, there are $3\times5$ grids (colored red)
  where a cut may present or absent. Each grid consist of $10\times16$
  rectangular graphene unit cells (colored green) and each rectangular
  graphene unit cell consists of four carbon atoms.}
\label{fig:fig1S}
\end{figure}

We used the Sandia-developed open source LAMMPS (Large-scale
Atomic/Molecular Massively Parallel Simulator) molecular dynamics (MD)
simulation code to model graphene~\cite{plimptonLAMMPS}. The
carbon-carbon interactions are described by the AIREBO
potential~\cite{stuart2000reactive}, which has been used previously to
study graphene kirigami~\cite{zenan-PRB-90-245437-2014}. The cutoffs
for the Lennard-Jones and the REBO term in AIREBO potential are chosen
to be 2~\AA~and 6.8~\AA, respectively. The graphene sheet of size
$\sim100\times200$~\AA~consisting 2400 (9600 carbon atoms) rectangular
graphene unit cells is shown in fig.~\ref{fig:fig1S}. In the 15 grids
(colored red), a cut of size $3\times16$ rectangular graphene unit
cells (colored green) is allowed to be present or absent. The graphene
kirigami were stretched by applying loads at the $+x$ and $-x$ edges
of the sheet. The atomic configurations were first relaxed by
conjugate gradient energy minimization with a tolerance of $10^{-7}$.
The graphene sheet was then relaxed at 4.2 K for 50 ps within the NVT
(fixed number of atoms $N$, volume $V$, and temperature $T$)
ensemble. Non-periodic conditions were applied in all three
directions. After the NVT relaxation, the edge regions were moved at a
strain rate of 0.01/ps, and the graphene sheet was stretched until
fracture.  This particular strain rate was chosen to save
computational time as it has been shown that the fracture strain and
fracture strength of graphene depend weakly on the strain rate,
especially for low temperature~\cite{zhao-JAP-108-064321-2010}. The 3D
stress was calculated as the stress parallel to the loading direction
times the simulation box size perpendicular to the plane and divided
by the graphene effective thickness of 3.7~\AA. Similar procedures
have been used for other MD, DFT simulations and
experiments~\cite{zhang-NatComm-5-3782-2014,
  hanakata-Nanoscale-8-458-2016, liu-PRB-76-064120-2007,
  hanakata-PRB-94-035304-2016, lee-science-321-385-2008}.

As the cuts are allowed to be present in any of the $3\times5$ grids there are
$2^{15}=32768$ possible cut configurations, which we will refer to as
the design space (or exploration space). Out of those, only 29791
configurations are not detached (no full cut along the $\hat{y}$
direction). Because the system is not periodic, translation symmetry
is broken. The reflection symmetry is not broken and thus only about
1/4 of the possible configurations need to be simulated via
MD. 
\section{Machine learning performance and molecular dynamics precision}
The root-mean-square error (RMSE) is given by, 
\begin{equation}
{\rm RMSE}=\sqrt{\frac{\sum^{m_{\rm test}}_{i=1}(y^{i}_{\rm test}-y^{i}_{\rm pred})^2}{m_{\rm test}}}
\end{equation}
where $m_{\rm test}$ is the number of test datasets, $y_{\rm test}$ are the
true values (obtained from MD) from the test dataset, and
$y_{\rm pred}$ are the predicted values from a model.

Because of thermal fluctuations (non-zero temperature), the obtained yield strain or the yield
stress of graphene from the MD simulations at 4.2 K will have
some variation, which we will refer to as the MD precision.  The MD
precision (irreducible error) for the yield strain $\eta^{\varepsilon}$ 
can be approximated as root-mean-square deviation (RMSD), 
\begin{equation}
 \eta^{\varepsilon}=\sqrt{\frac{\sum^{T}_{i=1}(\varepsilon^{\rm y}_{i}-\overline{\varepsilon^{\rm y}})^2}{T}}\,,
\end{equation}
where $T$ is the number of observations and
$\overline{\varepsilon^{\rm y}}$ is the average of
$\varepsilon^{\rm y}$ over $T$ observations which in this case are the
different initial velocities (different initial conditions).  The same
formula is used for the yield stress
$\eta^{\sigma}$. $\eta^{\varepsilon}$ is generally larger for systems
with more cuts. To save computational time we randomly choose a
configuration from systems having a cut density ranging from 0/15 to
12/15, calculate the RMSD from three different initial conditions, and
then sum the RMSD of each cut. Note that cut densities of 13/15--15/15
are not considered because the structure is fully detached. For yield
strain, the variability is more present at higher cut density and thus
we sum the RMSD from cut densities ranging from 5/15 to 12/15. In
addition to comparing RMSE and MD precision for evaluating the quality
of a model, we quantify the performance of a model using an $R^2$
score on the test dataset given by
\begin{equation}
R^2=1-\frac{\sum^{m_{\rm test}}_{i=1}|y^{i}_{\rm test}-y^{i}_{\rm pred}|^2}{\sum^{m_{\rm test}}_{i=1}|y^{i}_{\rm test}-\frac{1}{m_{\rm test}}\sum^{m_{\rm test}}_{i=1}y^{i}_{\rm pred}|^2}\,.
\end{equation}

\section{Machine learning}
\subsection{Hyperparameters for training}
We used open-source machine learning packages to build the machine
learning models. Specifically, we used TensorFlow
r1.8~\cite{tensorflow2015-whitepaper} for both the neural networks
(NN) and convolutional neural networks (CNN) model and
scikit-learn~\cite{scikit-learn} for the ordinary least square
regression model. The TensorFlow r1.8 was run on four CPUs and one
NVDIA Tesla K40m GPU card.

We will denote $h$ as the number of neurons in a hidden layer
$l$. In each layer the computation is given by
\begin{equation}
a^{(l+1)}=g(W^{(l)}a^{(l)}+b^{(l)}),
\end{equation}
where $g$ is the rectified linear unit (ReLU) activation function,
$a^{(l)}$, $b^{(l)}$, $W^{(l)}$ are the activations, bias, and the
weights in a layer $l$. In CNN, we denote `h' as the number of
fully-connected-layers (FCL) and `f' as the number of filters in a
convolutional layer. So for a CNN model with 16 filters in first
convolutional layer, 32 filters in second convolutional layer, 64
filters in third convolutional layer, and with a 64 FCL, we will
denote is as CNN-f16-f32-f64-h64. For both NN and CNN, a learning rate
of 0.0001 was used with a batch size of 200.  The number of maximum
epochs was set to 300.  A larger learning rate, e.g 0.001, or smaller
number total iteration number was found to have little effect on the
performance. For the search algorithm CNN-f16-f32-f64-h64, a learning
rate of 0.001 was used with a batch size of 100.

\subsection{Convolutional Neural Networks Architecture}
\begin{figure}
\includegraphics[width=10cm]{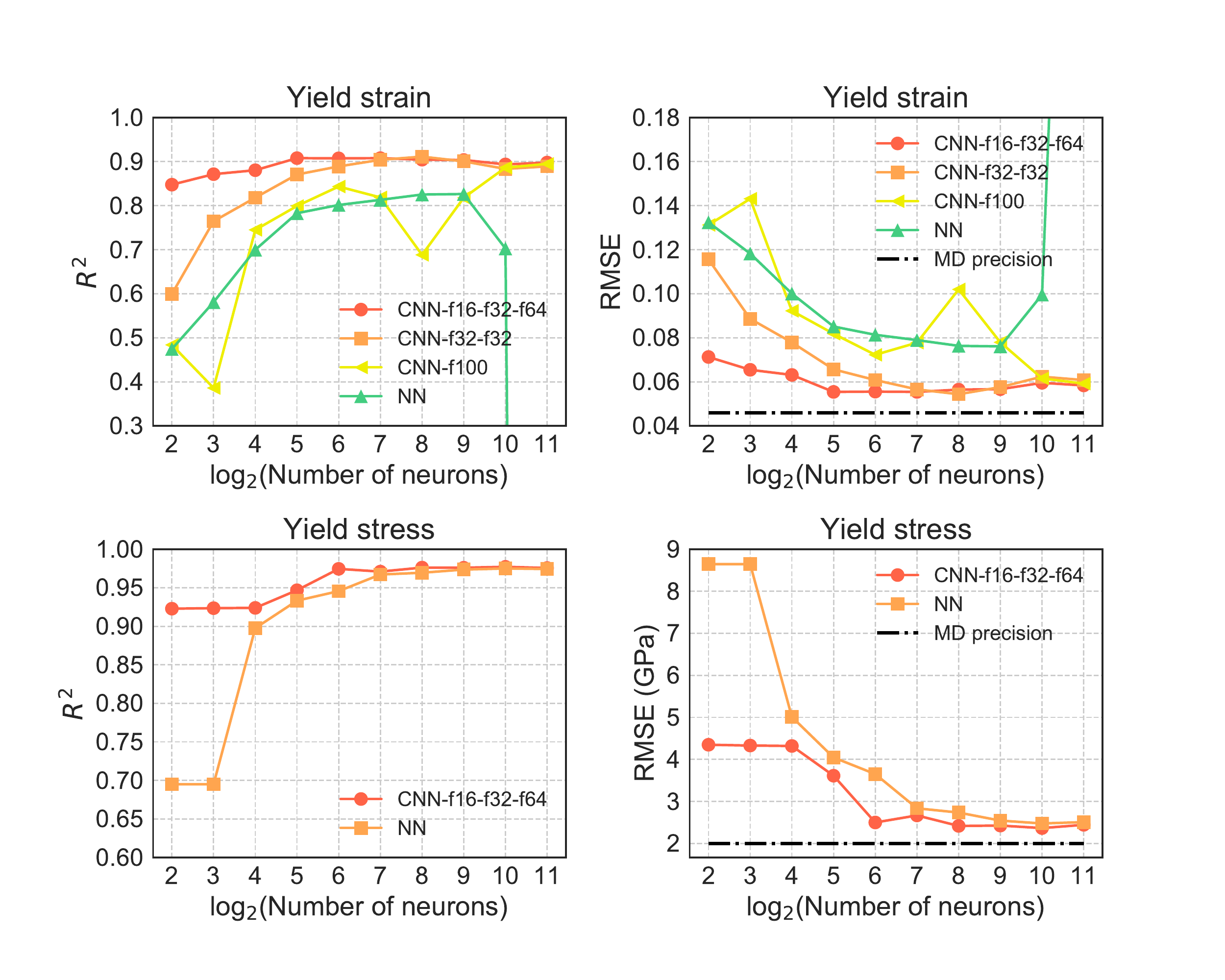}
\caption{$R^2$ and RMSE for yield strain (a, b) and yield stress (c,
  d) as a function of number of neurons for NN or size of
  fully-connected layer for CNN for different number of grids. In
  general, an increasing number of neurons increases the model
  accuracy. It can be seen that the CNN models outperform the NN
  models.}
\label{fig:fig2S}
\end{figure}

The input to the convolutional neural networks (CNN) was a fixed-size
$30\times80$ grey image (0/1 grids). We followed an architecture
similar to VGGNet~\cite{vggNet2014}. Specifically, the kernel size was
fixed at $3\times3$ with a stride of 1. Each convolutional layer was
followed by a ReLU function and a max-pooling layer of size $2\times2$
with a stride of 2.  The same padding was used after the first
convolutional layer to preserve the image size. We included one
fully-connected layer of size ranging from 4 to 2024. As we performed
regression, we did not include the ReLU function at the end of the
final layer.  The Adam optimizer was used to minimized the mean
squared error. 

\begin{figure}
\includegraphics[width=9cm]{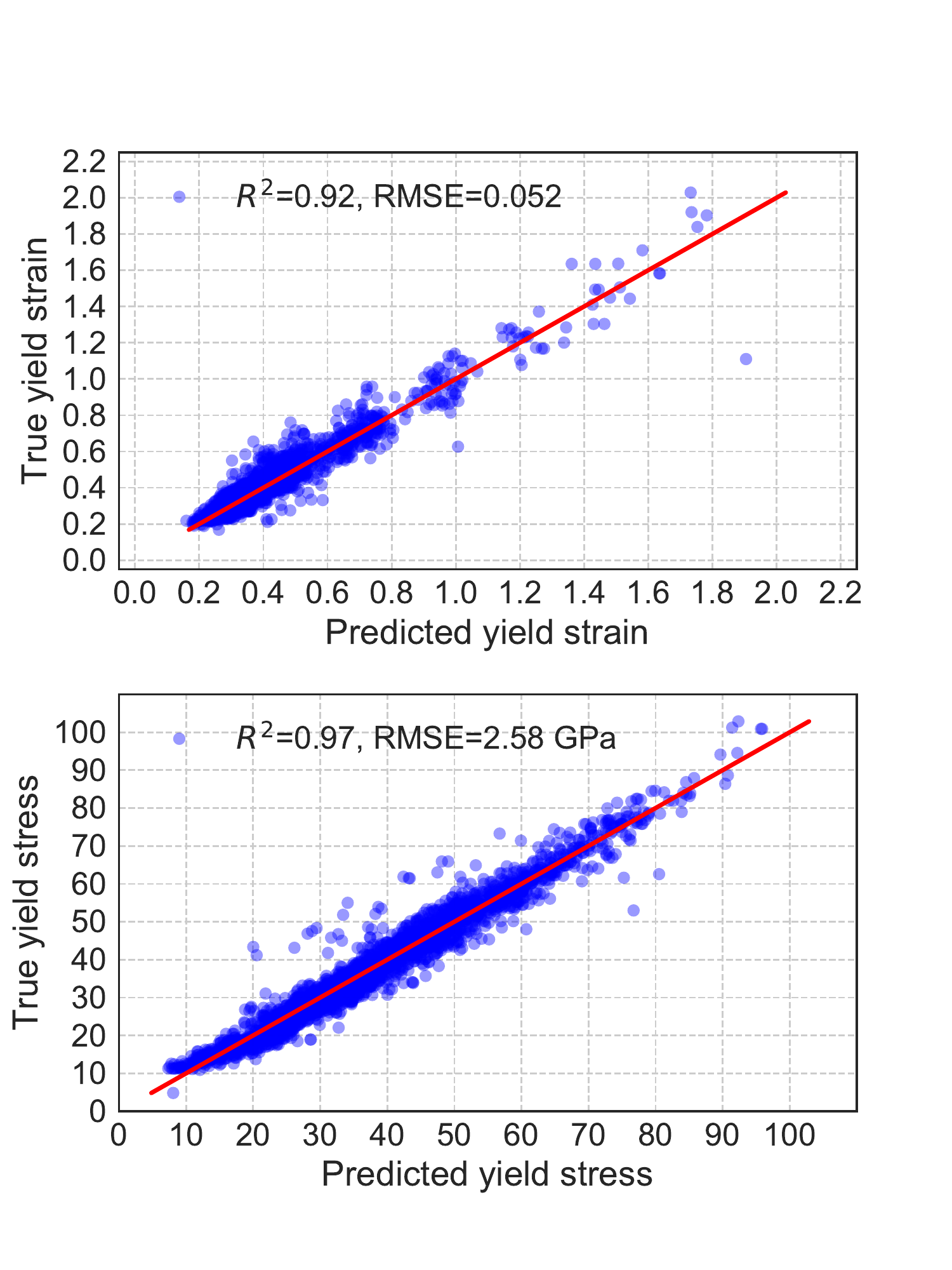}
\caption{Plot of true values (test dataset) as a function of predicted
  value for (a) yield strain and (b) yield stress. Here, the
  CNN-f16-f32-f64-h64 model was used. The red line represents $y=x$
  line.}
\label{fig:fig3S}
\end{figure}

In the supervised learning model, we split the 29,791 data samples
into 80\% for training, 10\% for validation, and 10\% for test
dataset. The validation dataset is used to find better architectures
("hyperparameter tuning", i.e. changing number of neurons or filters),
and the test dataset is used to assess performance. In each model, the
$R^2$ of the training dataset is slightly larger than the validation or
test dataset, indicating that there is no overfitting problem.  For
instance, for CNN-f16-f32-f64-h64 the $R^2$ are 0.93, 0.91, 0.92, on
training, validation, and test dataset respectively. The RMSE are
0.046, 0.056, 0.052 on training, validation, and test dataset
respectively. We found that the deep CNN architecture with increasing
number of filters from 16 to 64, similar to VGGNet
architecture~\cite{vggNet2014}, performed the best compared to the
wide CNN architectures or wide NNs. The performance comparison on the
validation dataset is shown in fig.~\ref{fig:fig2S}. In addition, we
also include performance comparison for yield stress shown in
fig.~\ref{fig:fig2S}(c) and (d).  Fig.~\ref{fig:fig3S} shows the
CNN-f16-f32-f64-h64 fitness in predicting yield strain and yield
stress on the test datasets.

\section{Rotations in kirigami}

\begin{figure}
\includegraphics[width=9cm]{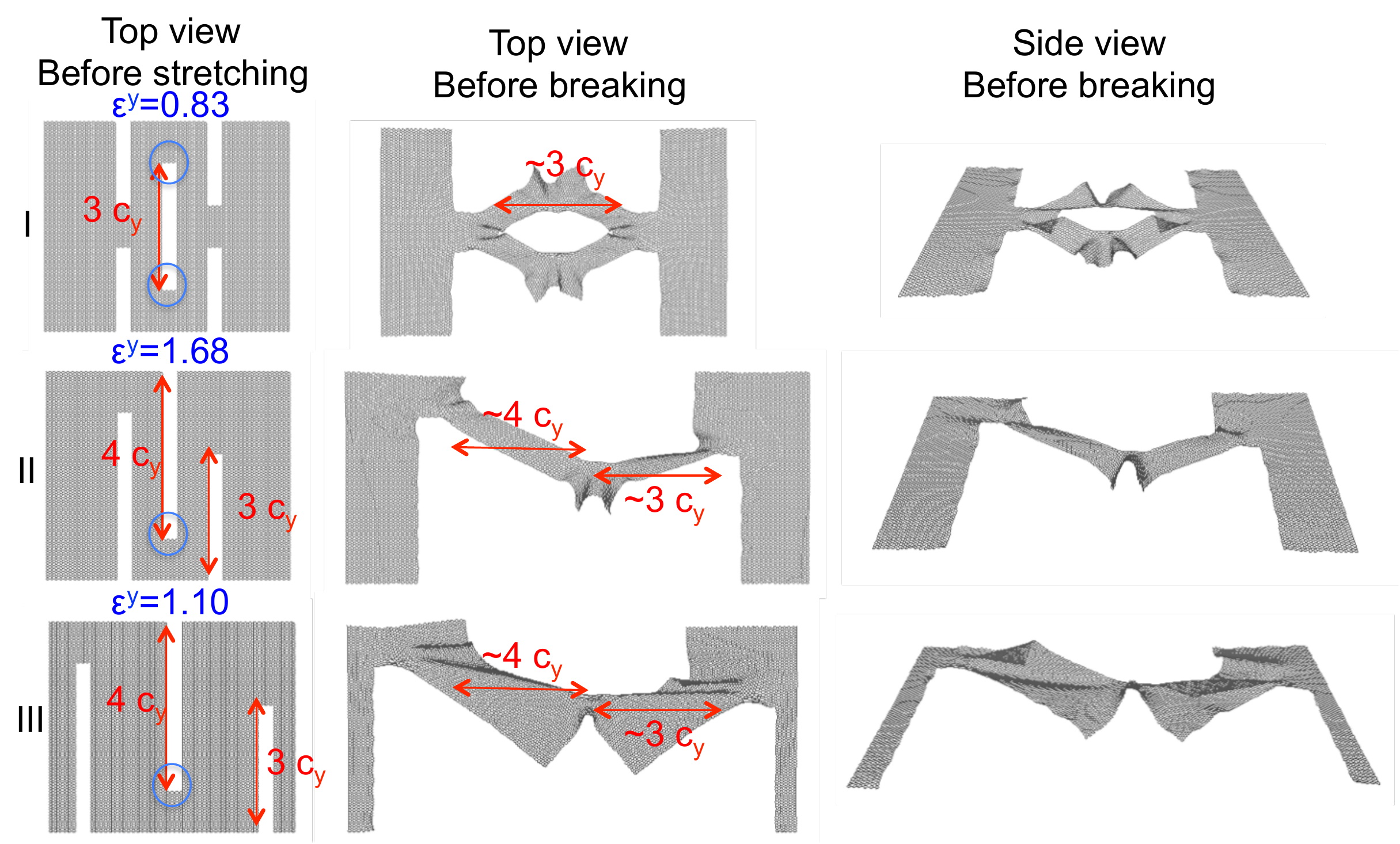}
\caption{Top and side views of three representative kirigamis.}
\label{fig:fig4S}
\end{figure}

The CNNs generate candidates with an extremely high yield strain
compared to the pristine (cut-free) graphene sheet. Here we
investigate the mechanisms enabling high stretchability, and focus on
rotation mechanisms in a few kirigami structures as shown in
fig.~\ref{fig:fig4S}. First we compare a kirigami with typical
centering cuts (fig.~\ref{fig:fig4S} I) and a kirigami with
alternating edge cuts (fig.~\ref{fig:fig4S} II). Let the dimension of
one cut be $c_x\times c_y$. We denote the ``meeting point'' in the
middle segment where two ribbons are connected as a node, which are
denoted by circles in fig.~\ref{fig:fig4S}. Assuming there is no bond
breaking, configuration I will have a maximum extension of
$\approx3c_x$ where the ribbons are connected by two nodes, whereas
configuration II will have a maximum extension $\approx7c_x$ as the
ribbon is connected by one node. This additional degree of freedom
(one fewer connecting node) enables configuration II to experience
significantly more elastic stretching as compared to configuration I.
As mentioned in the main text, the ML indeed found that the yield
strain of configuration II is almost {\it twice} as large that of
configuration I, as the alternating edge cut pattern effectively
transforms the 2D membrane to quasi-1D membrane.

An additional consideration is how the closeness of the alternating
edge cuts impacts the elastic stretchability.  As can be seen by
comparing configurations II and III in fig.~\ref{fig:fig4S}, close
packing of the alternating edge cuts allows increased stretchability.
This is because, as seen from the side view, the ribbons need to
rotate (twist) due to the applied tensile loading.  Because the cuts
are sparse in configuration III, the ribbons around the nodes are
thick. In configuration II, the ribbons are thinner, which improves
twisting and increases stretchability from 1.10 to 1.68.

\section{Linear model}
In addition to the NN and CNN models, we applied a simple linear model
to predict the yield strain and yield stress of the graphene
kirigami. We formulate the objective functions (yield stress and
strain) as
\begin{equation}
f(\pmb{x})=\beta_0+\sum_i\beta_ix_i+\sum_{i\leq j}\beta_{ij}x_ix_j+\sum_{i\leq j\leq k}\beta_{ijk}x_ix_jx_k+\dots\,.
\label{eq:eqS1}
\end{equation}
For $m$ samples, we can write Eq.~\ref{eq:eqS1} as a linear function
$f(\pmb{x})=\bf{X}^T\cdot\pmb{\beta}$, where
\begin{equation}
{\bf X}^T=\begin{pmatrix}
                1 & x^{(1)}_1 & \dots & x^{(1)}_n & x^{(1)}_1x^{(1)}_2 & \dots 
		\\
		\vdots & \ddots  & & & &
                \\
                1 & x^{(m)}_1 & \dots & x^{(m)}_n & x^{(\rm m)}_1x^{(m)}_2 & \dots 
	\end{pmatrix}\,,
\end{equation}
and
$\pmb{\beta}^T=(\beta_0, \beta_1, \dots, \beta_n, \beta_{12}, \dots)$.
In the machine learning language this is equivalent to applying
features transformation to the input vectors.  If the vector $\pmb{x}$
is binary, the infinite series reduces into a finite series with $2^n$
terms. Symmetries and locality will further reduce the number of
nonzero terms. For instance, in Ising and tight-binding models,
interactions are usually considered up to the first or second nearest
neighbors.

Since there is no theory that tells us the degree of the complexity,
we will increase the degree of polynomial until a reasonable
performance accuracy is achieved. We use the ordinary least squares
(OLS) regression to describe the yield stress as a function of
$\pmb{x}$. For this regression model we use the 15 long array that
distinguishes between cut and no cut in each grid as the input vector
$\pmb{x}$ and set the value of each component to be `1' for no-cut
and `-1' for cut. For yield strain, a polynomial of degree 3 gives
$R^2=0.76$ and ${\rm RMSE}=0.084$; for yield stress, a polynomial of 
degree 2 gives $R^2=0.93$ and root mean square error
${\rm RMSE}=4.1$ GPa.

\begin{figure}
\includegraphics[width=8cm]{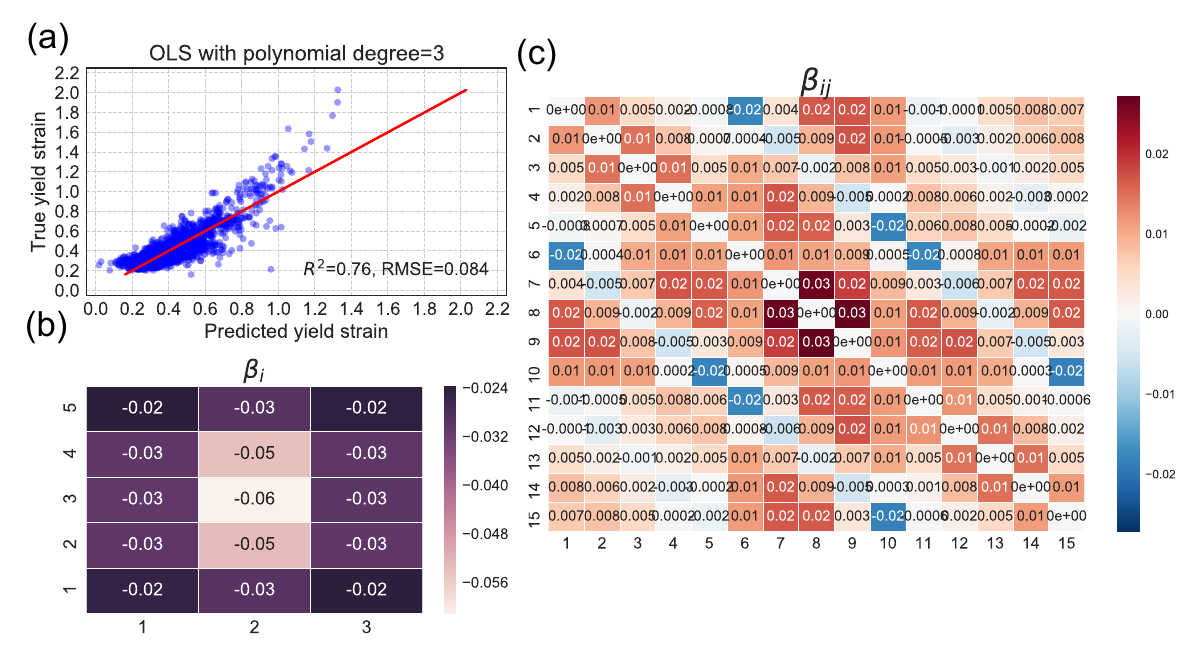}
\caption{(a) Linear plot of true values (test data) as a function of
  predicted value for yield strain. (b) Learned first order parameter
  $\beta_{i}$ plotted in 2D arrays to match with the real space
  positions for yield strain. Here (1,1) is $\beta_1$, (1, 2) is
  $\beta_2$ and so on. (c) Learned matrix second order parameters
  (coupling interactions) $\beta_{ij}$ for yield strain. The red line
  represents $y=x$ line.}
\label{fig:fig5S}
\end{figure}
To gain a physical understanding of how the kirigami should be
designed we plot the values of the parameters $\beta$. The first order
parameters $\beta_i$ are negative, suggesting that yield strains will
be higher when the materials have more cuts, as shown in
fig.~\ref{fig:fig5S}(b). The second order parameters $\beta_{ij}$
represent the pairwise `interaction coupling', and these give better
insights on how the kirigami should be designed in order to achieve
high yield strains. From fig.~\ref{fig:fig5S}(c), we see that the
values $\beta_{ij}$ are lowest (most negative) between two neighbors
along the $\vec{x}$ direction. On the other hand the coupling is
positive between two neighbors along $\vec{y}$. For instance
$\beta_{12}=0.01$ while $\beta_{16}=-0.02$.  The regression results
suggest that to achieve a high yield strain, cells (with a cut or no
cut) of same type should be placed long the $\hat{y}$, while the
opposite type should be placed right next to each other in the
$\hat{x}$. This means the kirigami should have a line of cuts in
$\hat{y}$ that alternate in the $\hat{x}$ direction. This resembles
mechanical springs with two different constants that are connected in
series.  Overall, the first order parameters $\beta_{i}$ tell us that
increasing cut density will result in higher strains; the second order
parameters $\beta_{ij}$ gives further design principles on how the
cuts should be arranged.

\begin{figure}
\includegraphics[width=8cm]{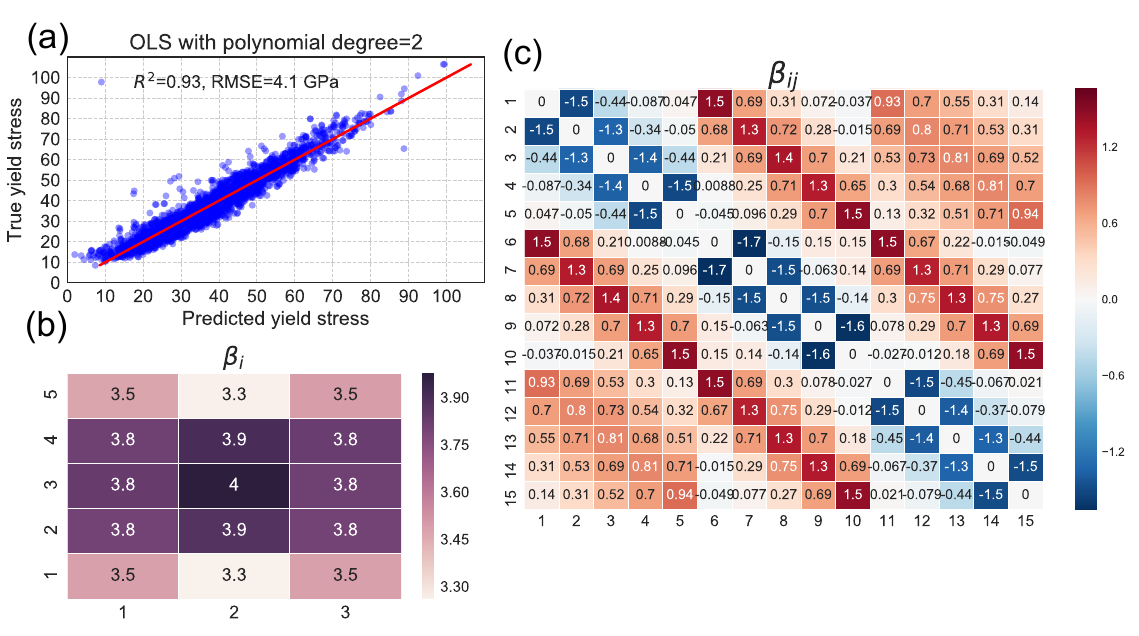}
\caption{(a) Linear plot of true values (test data) as a function of
  predicted value for yield stress. (b) Learned first order parameter
  $\beta_{i}$ plotted in 2D arrays to match with the real space
  positions for yield stress. Here (1,1) is $\beta_1$, (1, 2) is
  $\beta_2$ and so on. (c) Learned matrix second order parameters
  (coupling interactions) $\beta_{ij}$ for yield stress. The red line
  represents $y=x$ line.}
\label{fig:fig6S}
\end{figure}

As shown in fig.~\ref{fig:fig6S}(b), all first order parameters
$\beta_i$ for yield stress are positive, suggesting that introducing
cuts will lower the yield stress. From fig.~\ref{fig:fig6S}(c), we see
that the values are lowest (most negative) between two neighbors along
the $\vec{y}$ direction. On the other hand the coupling is positive
between two neighbors along $\vec{x}$. For instance $\beta_{12}=-1.5$
while $\beta_{16}=+1.5$. The regression results suggest that kirigami
should have arrays of cuts (or no cuts) in $\hat{x}$ that alternate in
the $\hat{y}$ direction. This resembles mechanical springs with two
different constants that are connected in parallel. In contrast to
{\it nearest neighbor} $\beta_{ij}$ of the yield strain, the nearest
neighbor for yield strain is positive $\beta_{ij}>0$ along $\hat{x}$
while $\beta_{ij}<0$ when the $j$ element is not in the same
$\hat{y}$.

We want to note that using series expansion works reasonably efficient
for {\it small} system; however, for finer grids (or larger systems),
the number of parameters will increase significantly as there will be
$(n+d)!/(n!d!)\sim n^d/d$ of $\beta$ terms, where $d$ is the
polynomial degree, $n$ is number of grids (features). Let us suppose
we use one rectangular graphene unit cell as the size of one
grid. Then, for a system size $\sim100\times200$~\AA, the input size
is $30\times80=2400$ ($10\times16$ rectangular graphene cells in each
coarse grid). At a polynomial degree of 3, we will need $\sim10^9$
parameters to fit.

In the series expansion approach, the 2D cut patterns are flattened
into a 1D-array and thus some of the local spatial information are
lost. Series expansion can be used to `recover' the information of
interactions between neighbors, but this approach becomes inefficient
when the number of cells becomes large. This series expansion approach
is computationally inefficient as we expect the interactions should be
local. In principle, one could do series expansions to the nearest
neighbors only; however, details of the potentials are not always
known. For these reasons, the CNN is a more appropriate and scalable
model as this deep learning is superior in recognizing edges in 2D
motifs as well as performing down-sampling, which is very suitable for
our problem. In the CNN, the 2D image (input) is convolved by a set of
learnable filters and this allows the model to learn 2D motifs. An
image passing through these filters then activates neurons which then
classify (or rank) the cutting patterns to good or bad designs. This
approach is more efficient than the series expansion approach as the
CNN model is built based on learned 2D local motifs.

\end{document}